%% file: main.tex
\documentclass[a4paper,twocolumn]{article}

\input{packages}

\input{commands}
\title{Evaluation of depth perception in crowded volumes}
\author{Žiga Lesar, Ciril Bohak, Matija Marolt}
\date{}

\begin{document}

\maketitle

\input{content/00-abstract}

\input{content/01-introduction}
\input{content/02-related-work}
\input{content/03-evaluation-setup}
\input{content/04-results}
\input{content/05-discussion}
\input{content/06-conclusion}

\bibliographystyle{plainurl}
\bibliography{bibliography}

\end{document}

%% file: packages.tex
\usepackage{amsmath}
\usepackage{graphicx}
\usepackage{hyperref}
\usepackage[nameinlink]{cleveref}

%% file: content/00-abstract.tex
\begin{abstract}
Depth perception in volumetric visualization plays a crucial role in the understanding and interpretation of volumetric data. Numerous visualization techniques, many of which rely on physically based optical effects, promise to improve depth perception but often do so without considering camera movement or the content of the volume. As a result, the findings from previous studies may not be directly applicable to crowded volumes, where a large number of contained structures disrupts spatial perception. Crowded volumes therefore require special analysis and visualization tools with sparsification capabilities. Interactivity is an integral part of visualizing and exploring crowded spaces, but has received little attention in previous studies. To address this gap, we conducted a study to assess the impact of different rendering techniques on depth perception in crowded volumes, with a particular focus on the effects of camera movement. The results show that depth perception considering camera motion depends much more on the content of the volume than on the chosen visualization technique. Furthermore, we found that traditional rendering techniques, which have often performed poorly in previous studies, showed comparable performance to physically based methods in our study.
\end{abstract}

%% file: content/01-introduction.tex
\section{Introduction}
\label{sec:introduction}

Volumetric data are ubiquitous in various scientific fields. They often originate from imaging techniques such as X-ray scans, computed tomography and magnetic resonance imaging, or from simulations in different domains. Effective visualization of such data is paramount for understanding its characteristics. Typically, this is achieved through volume rendering, where a transfer function is used to map raw volume samples to optical properties such as color and opacity. Depending on the chosen rendering technique, the rendering process may additionally account for transparency, illumination, occlusion, and projection to produce a 2D image. This is done to ensure adequate spatial perception of the viewer since mimicking or simulating realistic optical effects has been shown to enhance this aspect~\cite{Lindemann2011,Jonsson2014,Preim2016}. In essence, the primary objective of scientific volume visualization is to convey spatial information, including depth, distance, size, shape, and orientation as clearly as possible while providing users with the flexibility to interactively adjust certain visualization parameters, such as the transfer function and illumination setup.

Many visualization techniques promise to enhance spatial perception, a benefit that has undergone rigorous testing and validation in numerous studies. To ensure precision and repeatability, experiments in these studies are often conducted within highly controlled environments, with researchers making extensive efforts to isolate variables. As a result, such experiments typically involve only static images and a limited set of independent variables. Traditional volume rendering techniques tend to underperform in these constrained experiments due to their simplicity and lack of depth-enhancing features. An illustrative example of such a technique is the emission-absorption model~\cite{Max1995}. The widespread adoption of this rather simple technique in diverse domains attests to its ability to handle various tasks where visualization is needed.

Furthermore, previous experiments have focused on volumes representing recognizable structures, such as CT scans of the human body or angiographic images of blood vessels. These images predominantly feature familiar elements and can be rendered less visually dense by an appropriate transfer function. Both familiarity and visual density have been shown to profoundly affect spatial perception~\cite{Grosset2013,Englund2018}. This presents a major challenge when visualizing \emph{crowded volumes}, which are characterized by an overwhelming amount of mutually occluding structures. Conventional volume rendering techniques prove to be highly ineffective in this scenario due to the inherently complex structure of crowded volumes, making it imperative to use specialized tools~\cite{Weissenbock2014,LeMuzic2016,Lesar2023}. In this case, interactivity plays a much more important role, as users benefit from interactive grouping, colorization, and sparsification of the contained structures to lower their visual density. On the other hand, the substantial amount of occlusion observed in crowded volumes also has the potential to enhance spatial perception. According to Englund and Ropinski~\cite{Englund2016}, occlusion stands out as one of the strongest monocular depth cues, and the parallax effect induced by the camera movement can further amplify its significance. Therefore, when evaluating spatial perception in crowded volumes, it is essential to incorporate camera motion as an integral component of the visualization.

Previous studies have recognized the importance of camera motion in improving spatial perception. However, its evidently dominant positive effect has often led to its exclusion from evaluations. As a result, conventional rendering techniques have often underperformed due to the lack of essential depth cues. Furthermore, while studies have shown that the content of a volume has a significant impact on spatial perception, they have not considered crowded volumes or the degree of crowdedness as variables. In light of this, we conducted a study on depth perception in crowded volumes, in which we included camera motion to reflect a typical use case scenario. We employed two physically-based rendering techniques in addition to two traditional techniques for visualisation. We visualized three crowded volumes from different domains exhibiting varying degrees of crowdedness. Our goal was to find out how crowdedness and occlusion affect depth perception in crowded volumes. We also wanted to evaluate the performance of traditional rendering techniques under such conditions. Based on previous research and the above arguments, we developed the following hypotheses:
\begin{itemize}
 \item \textbf{(H1)} Crowdedness has a negative impact on depth perception in crowded volumes.
 \item \textbf{(H2)} It is easier to determine the depth in crowded volumes with larger instances.
 \item \textbf{(H3)} Traditional methods perform as well as modern methods in terms of depth perception.
\end{itemize}

The results of the study will help to identify the appropriate rendering techniques for visual inspection and analysis of crowded volumes.

The rest of the paper is organized as follows. \Cref{sec:related-work} lists the most important works and outcomes in volume visualization and perception. \Cref{sec:evaluation-setup} describes our experimental setup, the volumes and the techniques used. We report the results in \Cref{sec:results} and discuss their implications in \Cref{sec:discussion}. Finally, we present the conclusions in \Cref{sec:conclusion}.

%% file: content/02-related-work.tex
\section{Related work}
\label{sec:related-work}

We divide the related work into two subsections, covering visualization techniques and perception studies.

\subsection{Visualization techniques}

Visualization techniques can be broadly divided into non-realistic and physically-based techniques.

Non-realistic techniques include early optical models such as isosurface rendering~\cite{Levoy1988} and the emission-absorption model~\cite{Max1995}. Despite their age and limitations, both approaches are still well established in the visualization community. The non-realistic category of visualization techniques includes illustrative techniques and various smart visualization techniques~\cite{Viola2005}. Bruckner and Gröller~\cite{Bruckner2007} developed volumetric halos based on established medical illustration techniques, where partially occluded structures are darkened to create the illusion of depth. This idea has been later extended by Schott~et~al.~\cite{Schott2009} into a physically-based technique called directional occlusion shading (DOS). Several studies ranked DOS as one of the best techniques for enhancing depth perception~\cite{Lindemann2011,Preim2016,Diaz2017a}. In addition to illumination, researchers have explored leveraging other visual phenomena, such as depth of field~\cite{Grosset2013}, aerial perspective~\cite{Kersten2006}, and chromadepth~\cite{Ropinski2006,Bruckner2009,Diaz2010,KerstenOertel2013,Englund2016}, to improve depth perception. The motivation behind such a wide array of techniques that rely on chromadepth or pseudo-chromadepth is the observation that visualization techniques generally perform best when tasks can be reduced to simple comparisons~\cite{KerstenOertel2013}. In this manner, depth-related tasks are simplified to straightforward hue, brightness, or saturation comparisons. For a comprehensive overview of rendering techniques that are based on a close understanding of the human visual system to improve the images (so-called perception-driven techniques), we refer the reader to a survey by Weier~et~al.~\cite{Weier2017}.

The physically-based category includes techniques that aim to simulate light behavior in order to improve spatial perception, the reason being that the human visual system has evolved to interpret realistic illumination phenomena resulting from the underlying physics~\cite{Preim2016}. This category includes techniques such as path tracing~\cite{Lafortune1996}, diffusion simulation~\cite{Koerner2014}, combined volume and surface rendering~\cite{Kroes2012,Bohak2019,Smajdek2023}, spherical harmonic lighting~\cite{Kronander2012} and half-angle slicing~\cite{Kniss2002}. A comprehensive survey of illumination techniques in volume rendering is provided by Jönsson~et~al.~\cite{Jonsson2014}.

Both non-realistic and physically-based techniques benefit from visibility optimization techniques. This is especially true for crowded volumes. Chan~et~al.~\cite{Ming-YuenChan2009} presented transparency optimization to improve visibility of certain parts of a volume. Correa and Ma~\cite{Correa2011} introduced a similar idea based on visibility histograms. LeMuzic~et~al.~\cite{LeMuzic2016} extended the technique to crowded mesoscopic biological models by utilizing automatic sparsification. Lesar~et~al.~\cite{Lesar2023} adapted the visibility management concepts for the visualization of crowded volumes.

\subsection{Perception studies}

Perception in volume rendering has been the subject of extensive research. This is mainly due to the fact that perception is a complex process and is influenced by a wide range of factors, which are often hard to measure. Moreover, the importance of factors depends on the task at hand~\cite{Lindemann2011,KerstenOertel2013}. For example, no single visualization technique is optimal for both shape and depth perception~\cite{Preim2016}.

Human vision relies on various depth cues to deduce depth information from a scene. These can be categorized as either binocular, which are based on information from both eyes, and monocular, which can be observed with just one eye~\cite{Howard2002}. Computer graphics literature primarily focuses on monocular cues since binocular cues cannot be utilized when viewing images on a single 2D screen. Depth cues vary in their impact on depth perception, and they can be combined to enhance the effect~\cite{KerstenOertel2013,Hu2022}. For example, many visualization techniques rely on shadows and shading, and multiple studies have demonstrated their crucial role in depth perception~\cite{Wanger1992}. Interestingly, motion parallax and depth from motion are typically excluded from studies due to their substantial influence on depth perception, leaving only a handful of studies that include motion~\cite{Boucheny2009,KerstenOertel2013,Hu2022,Titov2022}. Díaz~et~al.~\cite{Diaz2017a} even state that it is disruptive. Some studies acknowledge the significance of motion as an important depth cue and suggest that further research is warranted in this area~\cite{Grosset2013,Englund2018}.

Boucheny~et~al.~\cite{Boucheny2009} noted that, ideally, one would see at a glance the whole volume and have a clear image in mind of the spatial layout. However, they also acknowledged that such representations of data are not natural because the human visual system usually experiences only solid surfaces. In their study, they demonstrate a significant preference for interpreting size as the main indicator of depth, even in the presence of conflicting information from other depth cues. Additionally, the strength of perspective has a notably positive impact on depth perception. Their study, however, had limitations due to the restricted range of rendering techniques and the use of very simplistic scenes.

Ropinski~et~al.~\cite{Ropinski2010} compared their shadow volume propagation technique to gradient-based shading and found significant improvements in speed and accuracy of depth perception. Lindemann and Ropinski~\cite{Lindemann2011} extended these results by conducting a larger user study in which they compared 7 state-of-the-art illumination models. They found that DOS~\cite{Schott2009} significantly outperformed other illumination models in ordinal depth perception and relative size perception tasks and performed very well in absolute depth perception tasks. In contrast to the previous study by Ropinski~et~al.~\cite{Ropinski2010}, participants expressed a preference for gradient-based shading over more advanced illumination models. The authors attribute this outcome to brighter and more colorful images produced by gradient-based shading. Interestingly, the study suggests that shadows may negatively affect size and depth perception.

Englund and Ropinski~\cite{Englund2016} assessed the perception of non-realistic volume rendering techniques and discovered that volumetric halos and depth-darkening techniques excel in absolute depth perception tasks. These techniques were also favored by the users. Ordinal depth perception, on the other hand, requires qualitative techniques, such as pseudo-chromadepth. The study was carried out using a crowdsourcing solution~\cite{Englund2016a}, which was shown to produce evaluation results comparable to those obtained with controlled experiments. The authors later expanded the study with feedback from three experts who stressed the significance of motion, clutter avoidance, and reduced transparency~\cite{Englund2018}.

Grosset~et~al.~\cite{Grosset2013} investigated the influence of depth-of-field on depth perception. Although they did not manage to validate any of their hypotheses, their study highlights substantial variations in depth perception across datasets. The research also indicates that familiarity has a notably positive effect on perception. Furthermore, anchoring proves to be beneficial, as objects suspended in mid-air pose challenges for spatial perception.

A great deal of research has already been carried out in the field of virtual reality (VR) and augmented reality (AR). Diaz~et~al.~\cite{Diaz2017} evaluated depth perception in AR with different settings for shadows, shading, perspective, dimensionality and texture. Their study showed that users' performance improved significantly with the use of shadows. In addition, their study showed that depth is consistently underestimated in AR. Díaz~et~al.~\cite{Diaz2017a} measured ordinal depth perception in stereoscopic displays of static images. They compared four techniques, with DOS performing the best and the emission-absorption model performing the worst. Interestingly, their study revealed no correlation between user accuracy and the on-screen distance between the estimation points. Hu~et~al.~\cite{Hu2022} investigated ordinal depth perception in VR on a neurological use case. They recognised the importance of shadows and motion parallax. However, their study found no significant differences in user performance when either of these depth cues was used. Heinrich~et~al.~\cite{Heinrich2019} conducted a study in a projective AR environment in which pseudo-chromadepth and support lines provided the best results.

In the field of medical visualization, considerable attention has been devoted to angiographic images. The study by Ropinski~et~al.~\cite{Ropinski2006} compared several visualization methods and revealed that angiography benefits from color coding depth information. This result was subsequently validated by Kersten-Oertel~et~al.~\cite{KerstenOertel2013}, who found that pseudo-chromadepth and aerial perspective resulted in the best user performance and were also favored by the users. Lesar~et~al.~\cite{Lesar2015a} assessed user preferences for various angiography visualization methods and found a strong inclination toward isosurface rendering with Phong shading and ambient occlusion. Drouin~et~al.~\cite{Drouin2020} explored dynamic visualization techniques, in which the visualization parameters are modified based on the motion of the VR controllers, applied in a surgical scenario. Their results unveiled significantly improved accuracy, albeit at a slower pace compared to static methods. Titov~et~al.~\cite{Titov2022} conducted a similar study in VR and found that the advantages of binocular disparity and motion parallax outweighed the benefits of dynamic techniques. Kreiser~et~al.~\cite{Kreiser2021} explored the use of empty space surrounding vessels to convey depth information and observed substantial improvements in ordinal depth perception.

In summary, although there are numerous volume rendering techniques that enhance depth perception, selecting a specific technique largely depends on the task at hand. Motion has received little attention in previous studies, and to the best of our knowledge, no study has yet explored depth perception in crowded volumes.

%% file: content/03-evaluation-setup.tex
\section{Experiment setup}
\label{sec:evaluation-setup}

The aim of the study was to investigate the effect of crowdedness, occlusion and camera motion on depth perception in crowded volumes using different rendering techniques.

The study was conducted in the form of an online crowdsourced survey whose target group comprises the general population. We did not focus exclusively on people who have experience with volume rendering, visualization or 3D modeling. Participants were asked to estimate the relative depth between structures after being shown a short video of a 3D volume. The evaluation consisted of 128 test cases in which four volumes were rendered using four different rendering methods and shown from eight different perspectives. The videos used in the test cases have been made publicly available~\cite{Lesar2024}.

\subsection{Evaluation interface}

\begin{figure}
    \centering
    \includegraphics[width=\linewidth]{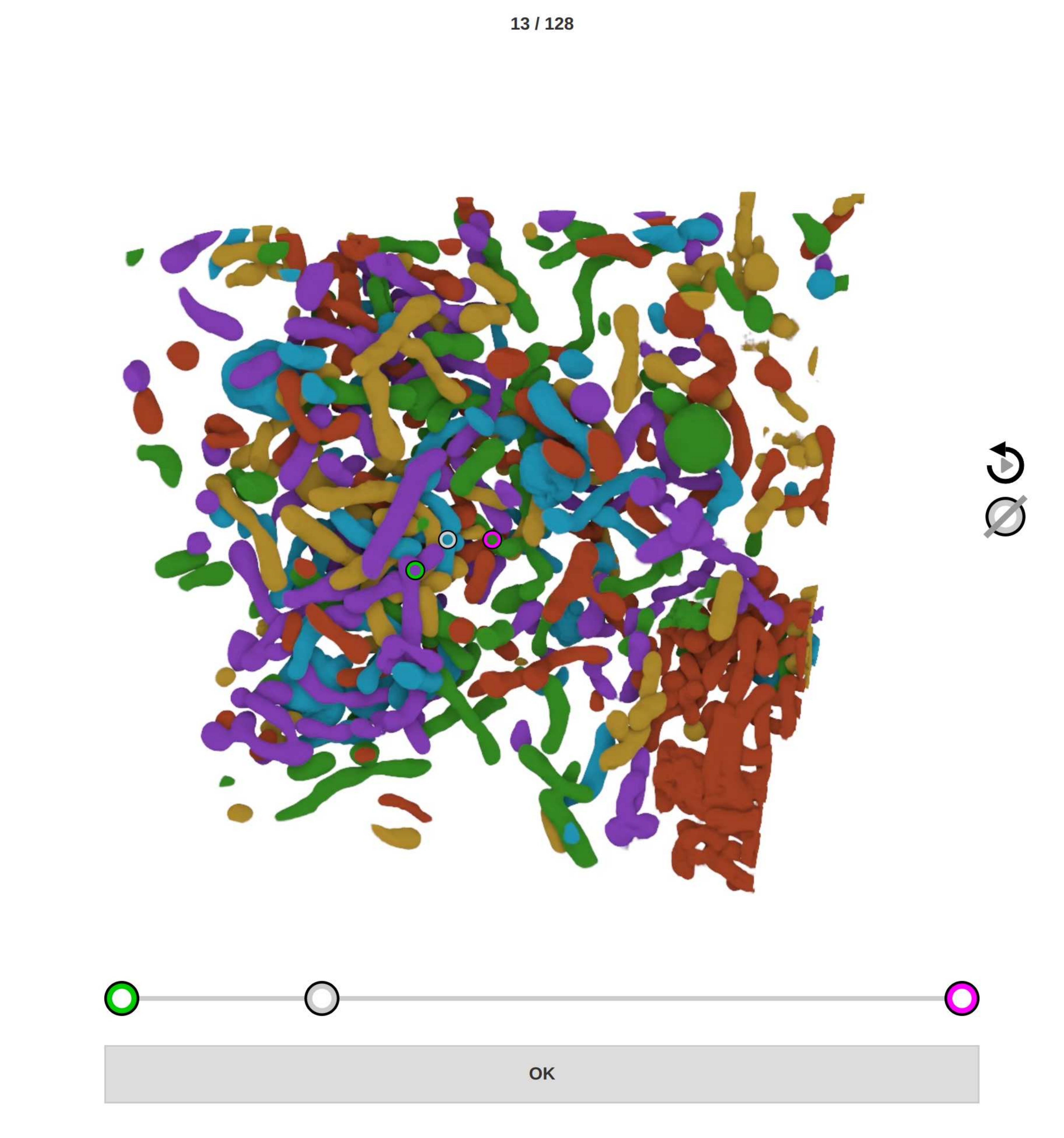}
    \caption{A screenshot of the user interface during the evaluation. The slider below the rendering was used to estimate the depth of the grey estimation point between the green and purple bounding points.}
    \label{fig:eval-screenshot}
\end{figure}

Each evaluation session began by showing participants background information about the study and the crowded volumes. Participants were asked to enter basic demographic data: gender, age, occupation, and experience with 3D environments (VR, gaming, and 3D modelling). They were then given instructions on how to complete each task. Specifically, for each task, they were shown a 5-second video of a volume in which the camera performed a subtle circular motion while keeping the volume in focus (see \Cref{sec:views} for details). When the video stopped playing, three points lying on the structures in the volume were shown, and participants were instructed to estimate the relative depth of the \emph{estimation point} between the two \emph{bounding points} using a slider (see \Cref{fig:eval-screenshot}). The points were marked with color-coded circles: the point closest to the camera was marked with a green circle, the point furthest away from the camera with a purple circle and the estimation point with a grey circle. The colors were chosen to be complementary and approximately equally bright, but at the same time as neutral as possible to avoid conveying unintended information to the participants. In particular, we tried to avoid the combination of green and red, which could be misinterpreted as ``right'' and ``wrong''. Participants were allowed to replay a video once to get a better sense of the spatial relationships between the structures in the volume. During video playback, the three points were hidden to avoid shifting the participants' focus toward the parallax effect rather than the actual rendering, as it is possible to estimate the relative depth of the three points based solely on their apparent movement due to the parallax effect.

\begin{figure}
    \centering
    \includegraphics[width=\linewidth]{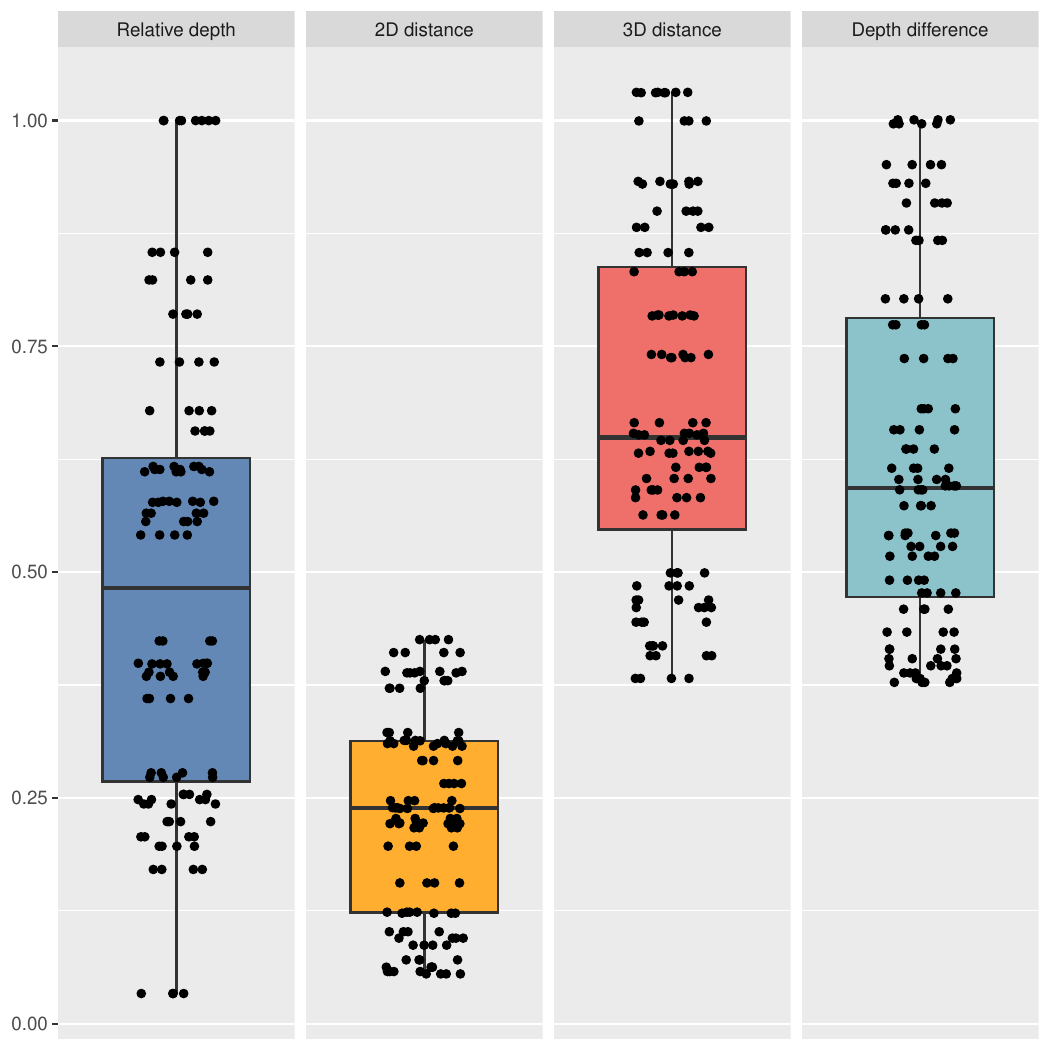}
    \caption{Distribution of distances and depths between evaluation points among test cases: relative depth of the estimation point, on-screen (2D) distance between the bounding points, volumetric (3D) distance between the bounding points, and depth difference between the bounding points. The on-screen distance has been computed in normalized device coordinates, and the volumetric distance has been computed such that the volume was a cube with a unit side length.}
    \label{fig:distances-boxplot}
\end{figure}

The positions of the points were manually chosen to adhere as closely as possible to the following criteria:
\begin{itemize}
    \item the relative depth of the estimation point should be approximately uniformly distributed,
    \item the depths of the bounding points should be approximately uniformly distributed inside the bounds of the volume,
    \item the points should be approximately uniformly distributed across the display, and
    \item the points should not be occluded by structures in the volume when the video stops.
\end{itemize}
The statistics for the above metrics across all test cases are shown in \Cref{fig:distances-boxplot}.

\subsection{Volumes}

\begin{figure}
    \centering
    \includegraphics[width=0.5\linewidth]{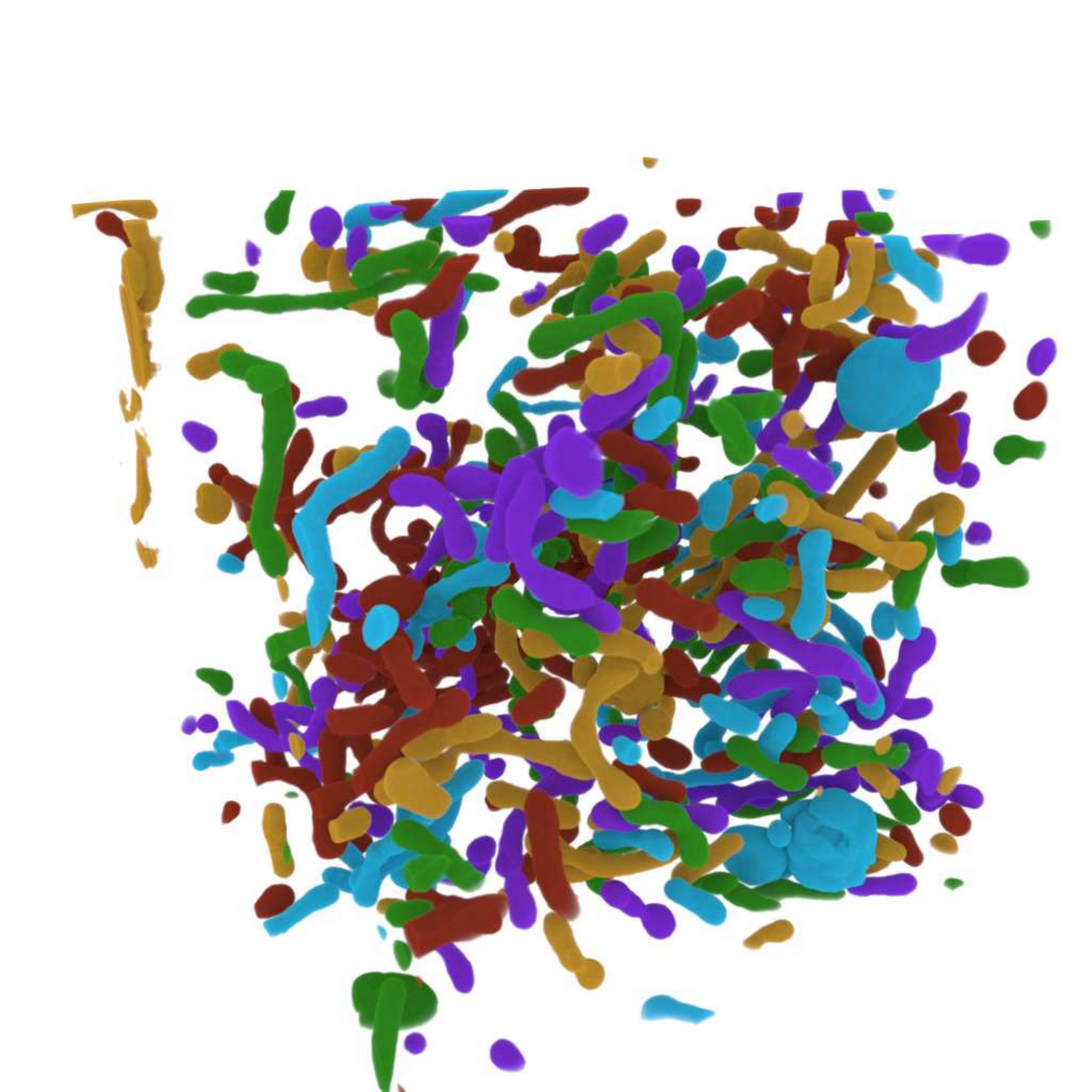}%
    \includegraphics[width=0.5\linewidth]{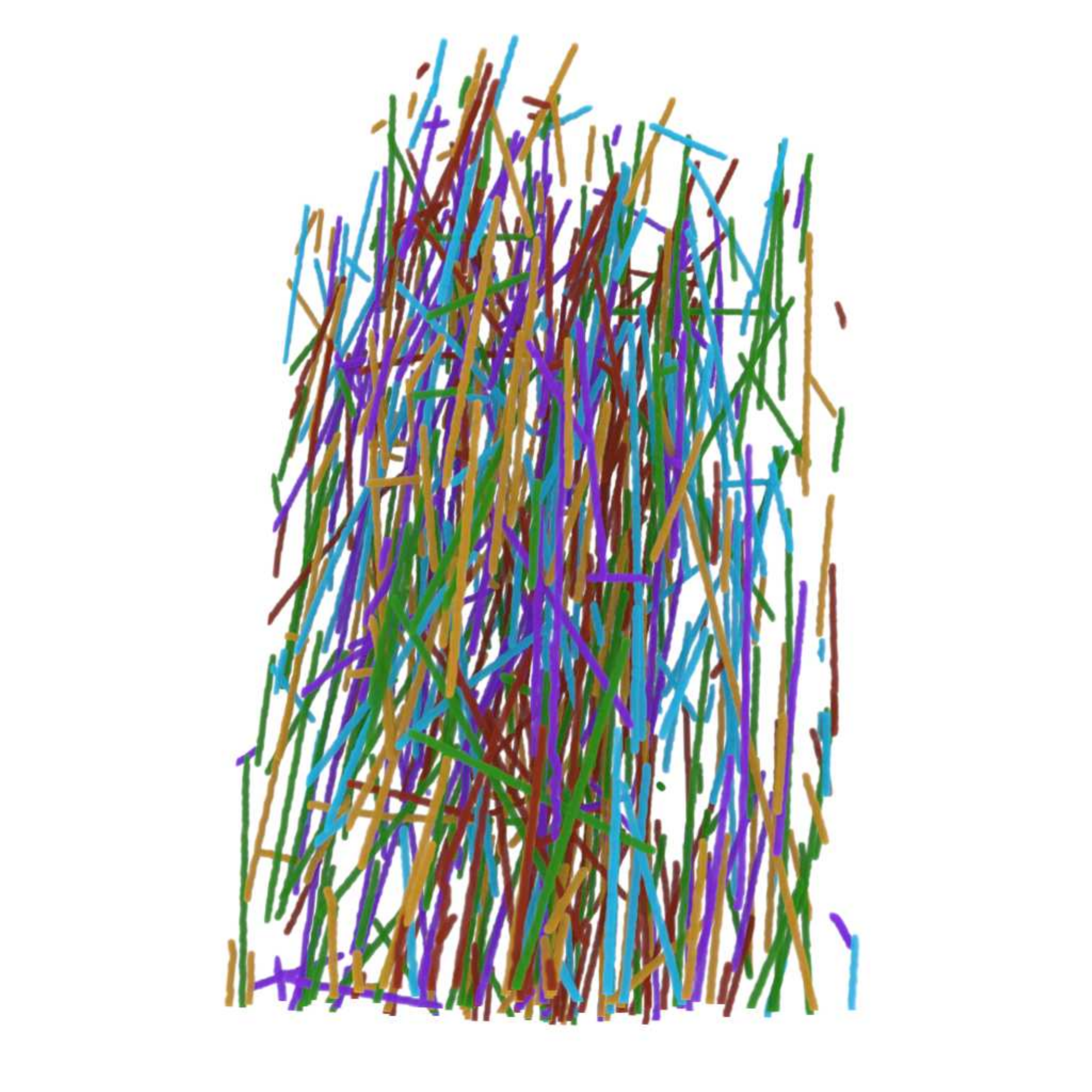}
    \includegraphics[width=0.5\linewidth]{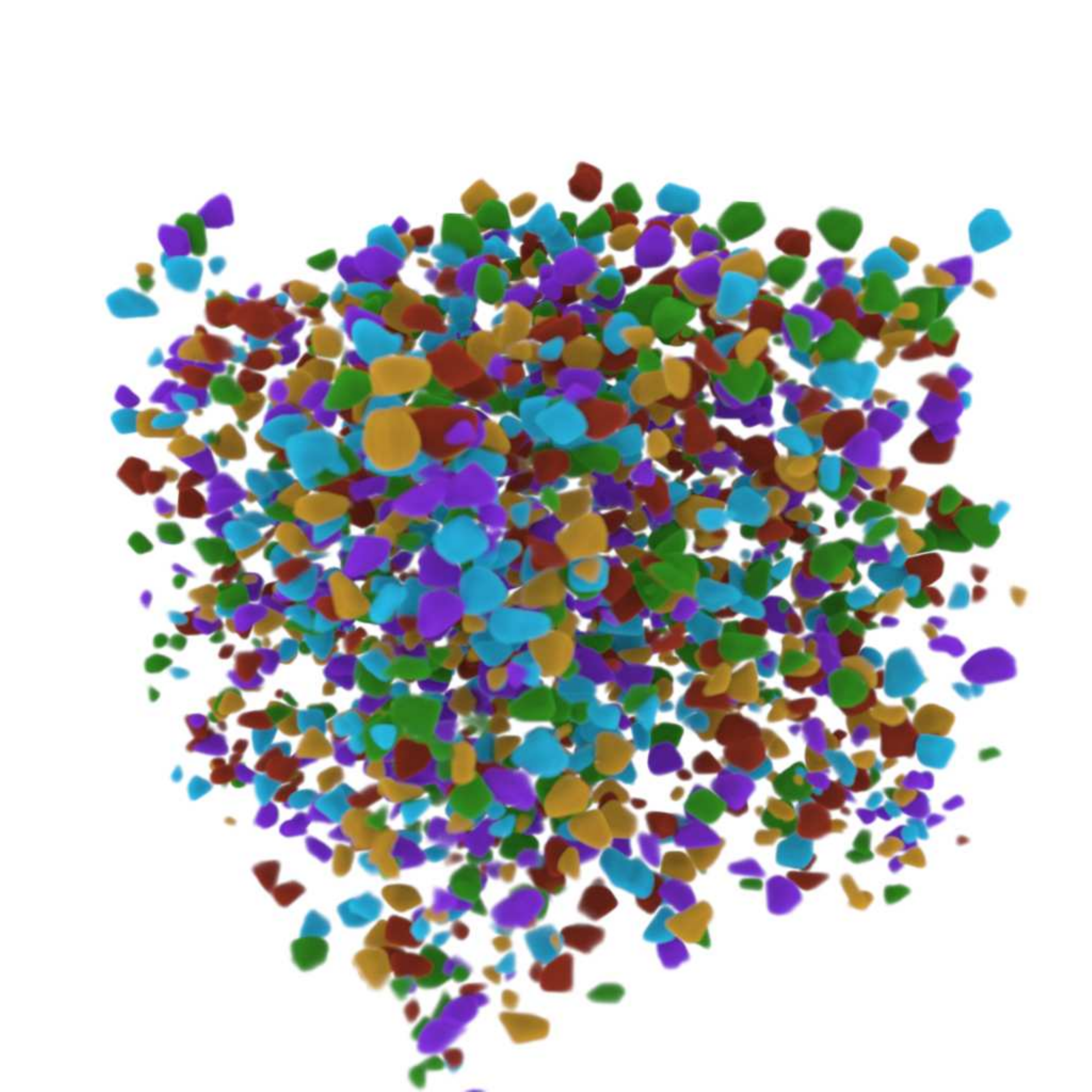}%
    \includegraphics[width=0.5\linewidth]{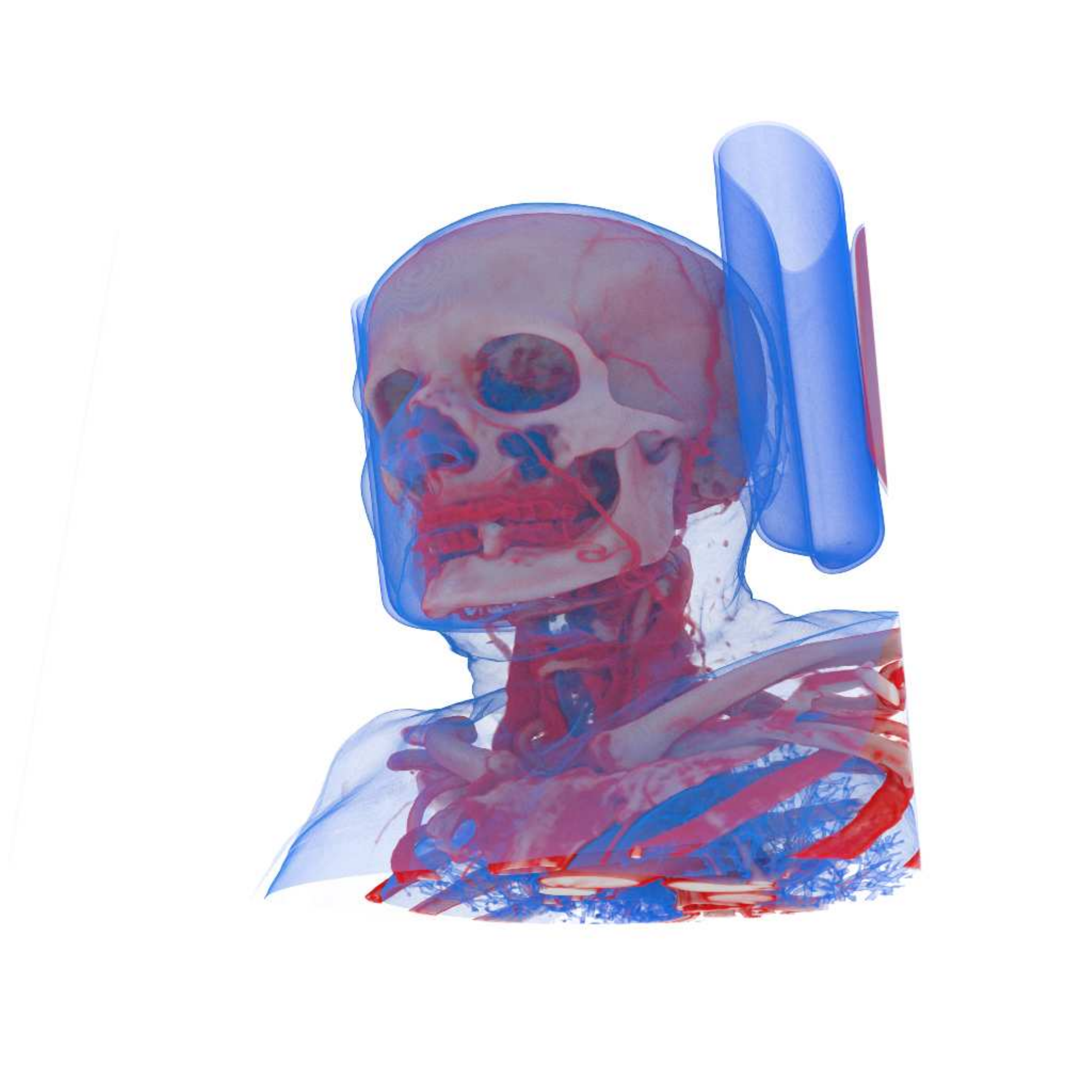}%
    \caption{Test volumes: mitos (top left), fibers (top right), pebbles (bottom left), and manix (bottom right). The images were rendered with volumetric path tracing.}
    \label{fig:volumes}
\end{figure}

We included four volumes in the study, three of which were crowded and the remaining one served as a control volume. The volumes are shown in \Cref{fig:volumes}. They come from different domains and their crowdedness varies considerably.

The first volume is from material science~\cite{Weissenbock2014} and represents an X-ray computed tomography image of a fiber-reinforced polymer that belongs to a family of materials that are in high demand in the automotive and aerospace industries due to their strength, durability, elasticity and light weight. Experts working with such materials analyze their properties by examining the length, orientation and distribution of the fibers. Due to the high fiber density, the tasks associated with the analysis of fiber-reinforced polymer materials require the use of special tools such as Fiber Scout~\cite{Weissenbock2014} or Volume Conductor~\cite{Lesar2023}. The volume we used for the user study measures $400 \times 401 \times 800$ voxels and contains 3828 instances of glass fibers. We will refer to this volume as \emph{fibers}.

The second volume comes from the biological domain. It is a $1280 \times 1024 \times 1024$ focused ion beam scanning electron microscopy (FIB-SEM) image capturing a segment of a mouse's bladder cell. The image was segmented to highlight the different intracellular organelles~\cite{ZerovnikMekuc2020,ZerovnikMekuc2022}. In addition to the numerous fusiform vesicles, Golgi apparatus, endosomes and lysosomes, there are 3051 mitochondria in the volume. Analyzing their distribution is of crucial importance for understanding the biological processes within a cell. We will refer to this volume as \emph{mitos}.

The third volume is synthetic and was created to contain a large number of instances. The volume of size $512 \times 512 \times 512$ contains 22670 instances of small convex polytopes resembling pebbles. The instances were physically simulated to achieve the densest possible packing within the contained volume. We will refer to this volume as \emph{pebbles}.

The fourth volume is a control volume that was included in the study to allow comparison with previous studies. The volume is a $512 \times 512 \times 460$ CT image of the upper torso of a human\footnote{Manix dataset from \url{http://www.osirix-viewer.com/datasets/}}. We will refer to this volume as \emph{manix}.

To reduce crowdedness to a level where the interior of the volume is at least partially visible, the pebbles were uniformly sparsified to 10 \% density, and the fibers were uniformly sparsified to 20 \% density. To represent a typical usage scenario, where the instances are grouped according to some criterion, the instances in all crowded volumes were divided into 5 groups of approximately equal size, and the colors have been assigned to the groups according to the golden ratio sequence~\cite{Schretter2012}, which guarantees an approximately uniform distribution of hues.

\subsection{Rendering methods}

\begin{figure}
    \centering
    \includegraphics[width=0.5\linewidth]{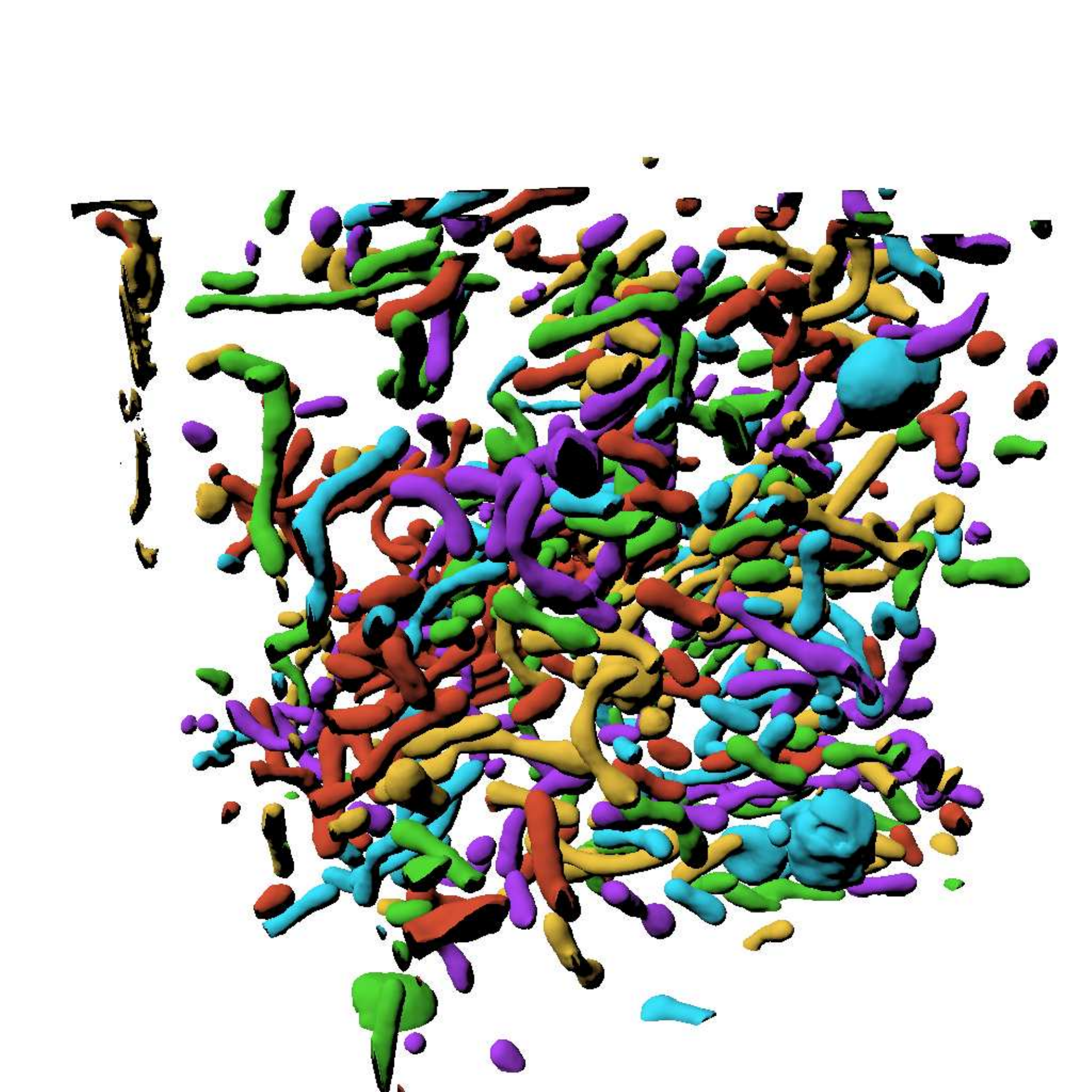}%
    \includegraphics[width=0.5\linewidth]{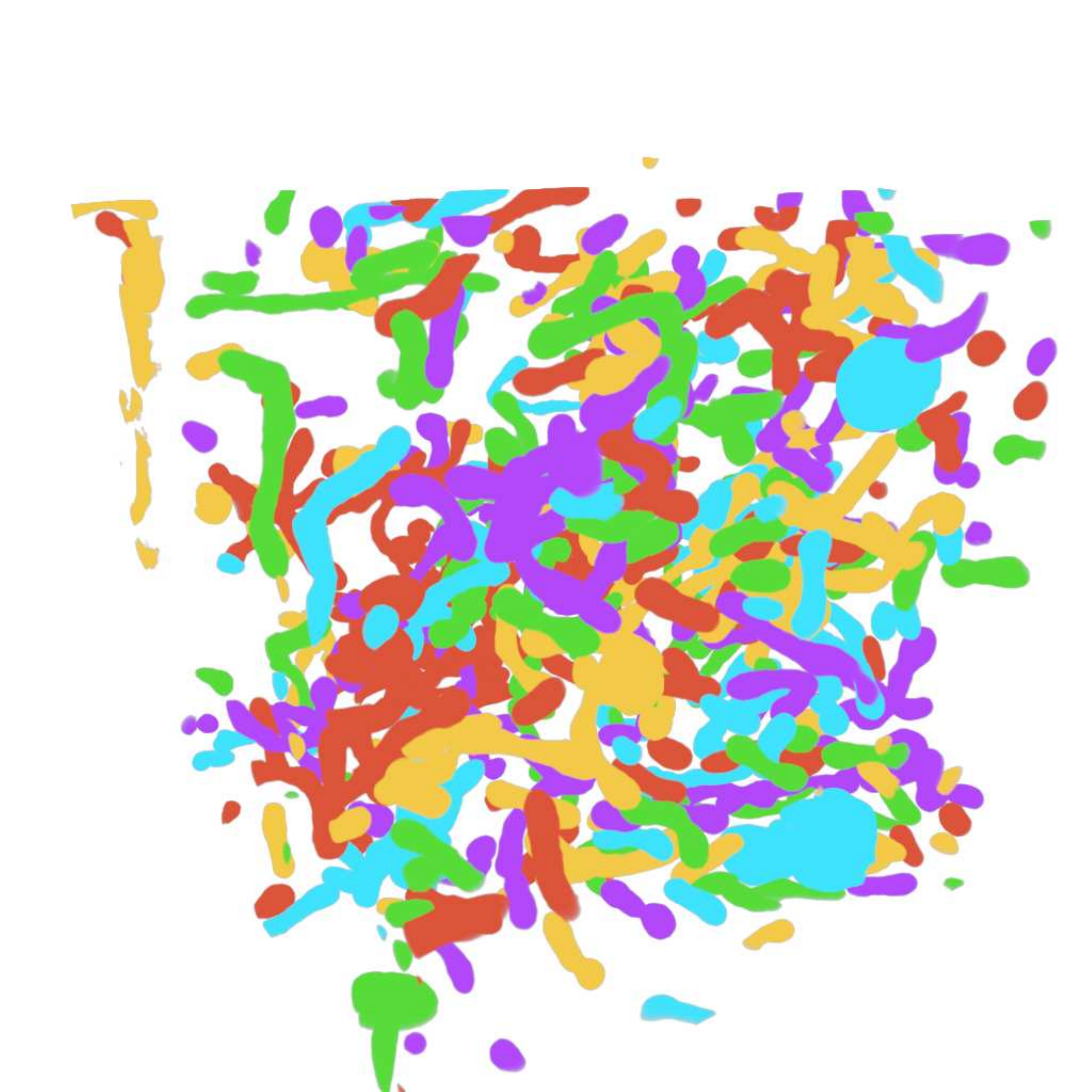}
    \includegraphics[width=0.5\linewidth]{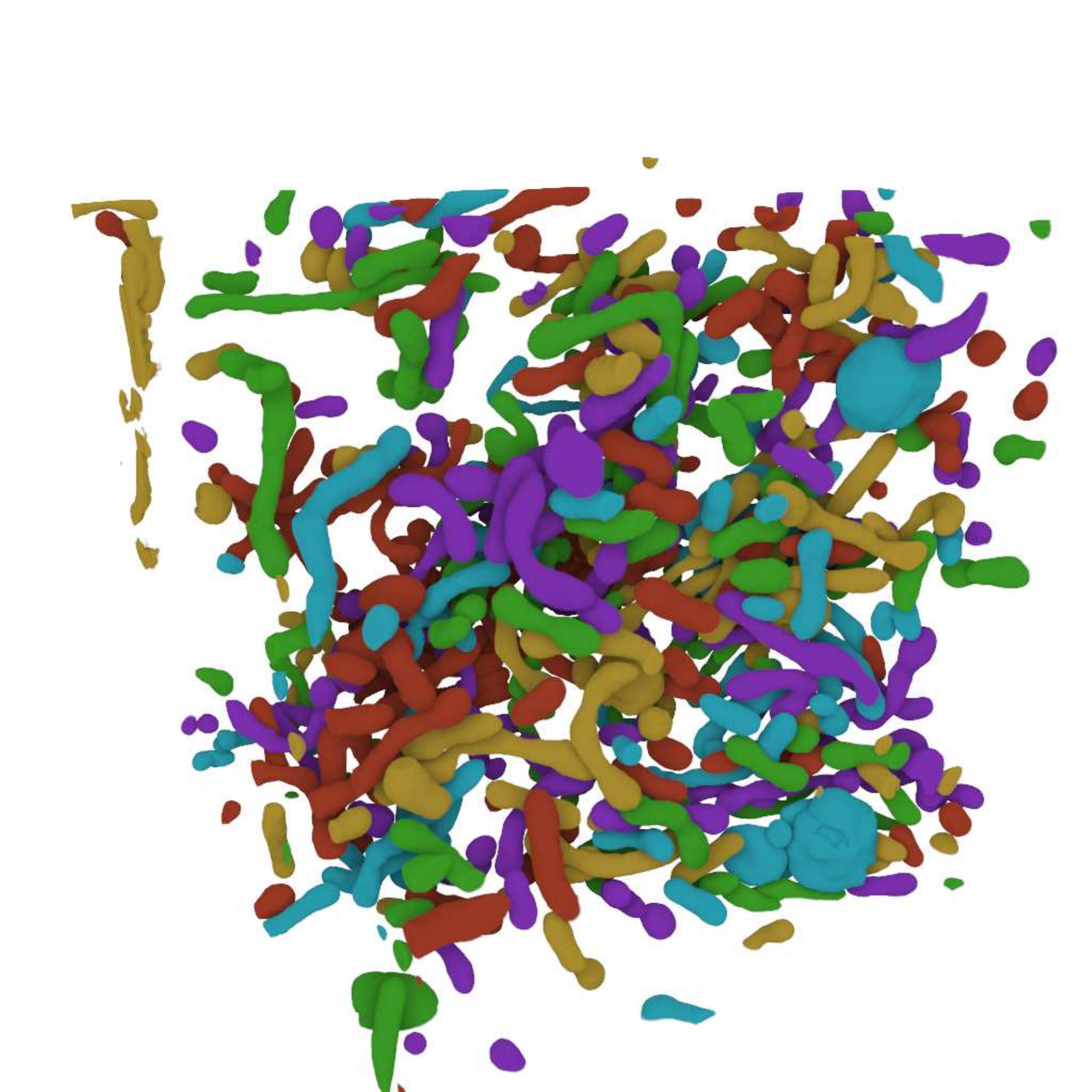}%
    \includegraphics[width=0.5\linewidth]{figures/renders/test0027.pdf}%
    \caption{The rendering methods used in the experiment: ISO (top left), EAM (top right), DOS (bottom left), and VPT (bottom right).}
    \label{fig:methods}
\end{figure}

For the visualization we used 4 widely used volume rendering methods, namely the Emission Absorption Model (EAM)~\cite{Max1995}, Directional Occlusion Shading (DOS)~\cite{Schott2009}, Isosurface Raycasting (ISO)~\cite{Levoy1988} and Volumetric Path Tracing (VPT)~\cite{Lafortune1996}. \Cref{fig:methods} shows a visual comparison of the four methods. In all renderings, we used a white illuminant. The isosurfaces were shaded using the Disney's physically-based model~\cite{McAuley2012} with a directional light and a fully rough material. For the directional occlusion model, an illumination cone with an apex angle of 60° was used. For volumetric path tracing, the multiple scattering model was used with a slightly backscattering medium modeled with a Henyey-Greenstein phase function~\cite{Toublanc1996} with the anisotropicity parameter of -0.5. The volume was illuminated with an infinite environmental isotropic white light.

These methods were implemented on top of the platform-independent volume rendering framework VPT~\cite{Lesar2018} and extended by the Volume Conductor interface~\cite{Lesar2023} with the capabilities to render and sparsify crowded volumes. No additional depth-enhancing methods, such as aerial perspective, pseudo-chromadepth, or volumetric halos were used to improve the renderings. This decision was made to limit the number of independent variables examined in the study.

\subsection{Views}
\label{sec:views}

To ensure comparability between different rendering methods, we generated 8 camera views that captured the data as uniformly as possible from different angles. We used spherical Fibonacci mapping~\cite{Keinert2015} to generate the camera positions on the bounding sphere surrounding the volume and aimed the camera at the center of the volume so that the up vectors were aligned in all configurations. The vertical field of view of the camera was set to 90°.

The camera performed a small circular motion around the base of a cone with the apex at the center of the volume and an apex angle of approximately 11°. The apex angle was chosen so that the parallax effect was sufficiently clear but not excessive.

%% file: content/04-results.tex
\section{Results}
\label{sec:results}

\begin{figure}
    \centering
    \includegraphics[width=\linewidth]{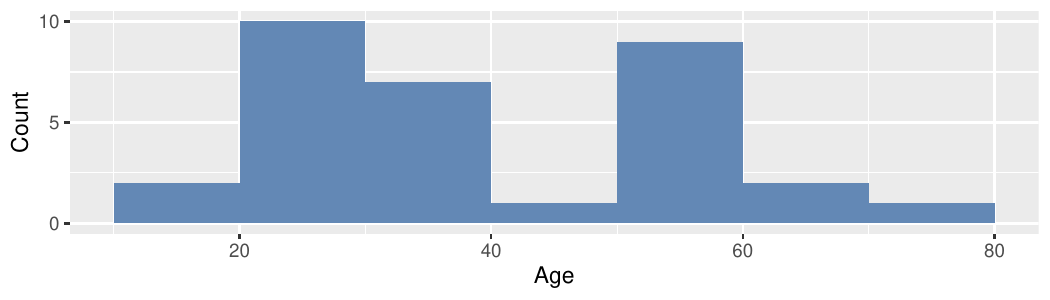}
    \caption{Distribution of ages among study {par\-ti\-ci\-pants}.}
    \label{fig:demographic-age}
\end{figure}

\begin{figure}
    \centering
    \includegraphics[width=\linewidth]{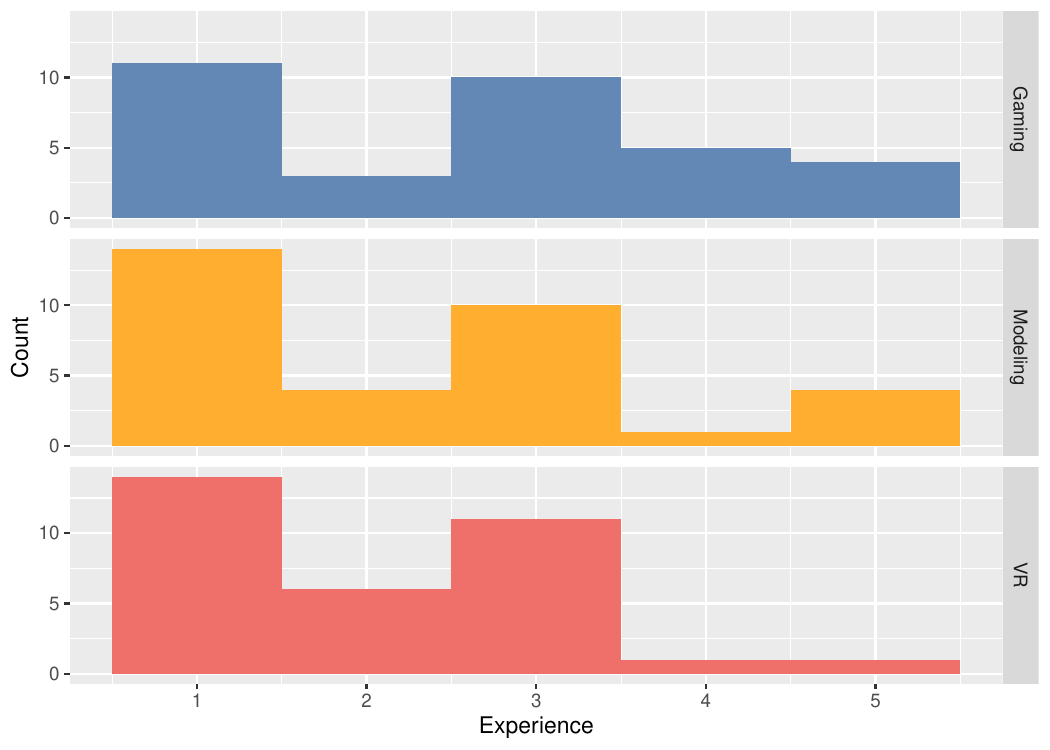}
    \caption{Distribution of the participants' prior experience with gaming, 3D modeling, and VR.}
    \label{fig:experience}
\end{figure}

The experiment was online for 46 days. During that time, we received 4340 depth estimates, which have been made publicly available~\cite{Lesar2024}. The estimates were provided by 47 participants, 29 of whom completed the entire set of 128 depth estimation trials. 33 participants reported demographic data. 21 were women, and 12 were male. Their ages ranged from 14 to 74, with a mean age of around 38 (see \Cref{fig:demographic-age}). The participants reported their prior experience with 3D environments (games, 3D modeling, and VR) on a 5-point Likert scale ranging from 1 (no experience) to 5 (a lot of experience). The average gaming experience reported by the participants was 2.64, with a standard deviation of 1.41. 3D modeling experience was reported with a mean of 2.30 and a standard deviation of 1.38. VR experience had a mean of 2.06 and a standard deviation of 1.09. Histograms illustrating the distributions of these experiences are shown in \Cref{fig:experience}.

To evaluate depth perception, we measured the difference between the correct and estimated relative depths. The estimated depth is given by the participants using a slider (see \Cref{fig:eval-screenshot}), while the correct relative depth is calculated using a depth buffer associated with the rendering. Both numbers fall within the range of 0 to 1 due to the nature of their acquisition. To test our hypotheses, we relied on the mean absolute difference (MAD) between the correct and estimated depths. In theory, if the correct and estimated depths are independent and uniformly distributed on the unit interval, then the expected MAD is $1/3$. We use this number as a reference for interpreting the measured data.

Since the experiment was uncontrolled, the participants could pause the evaluation or refresh the page at any time during the experiment. As a result, we could not reliably measure the response time, which is often analyzed as an important performance metric in controlled experiments.

\begin{figure}
    \centering
    \includegraphics[width=\linewidth]{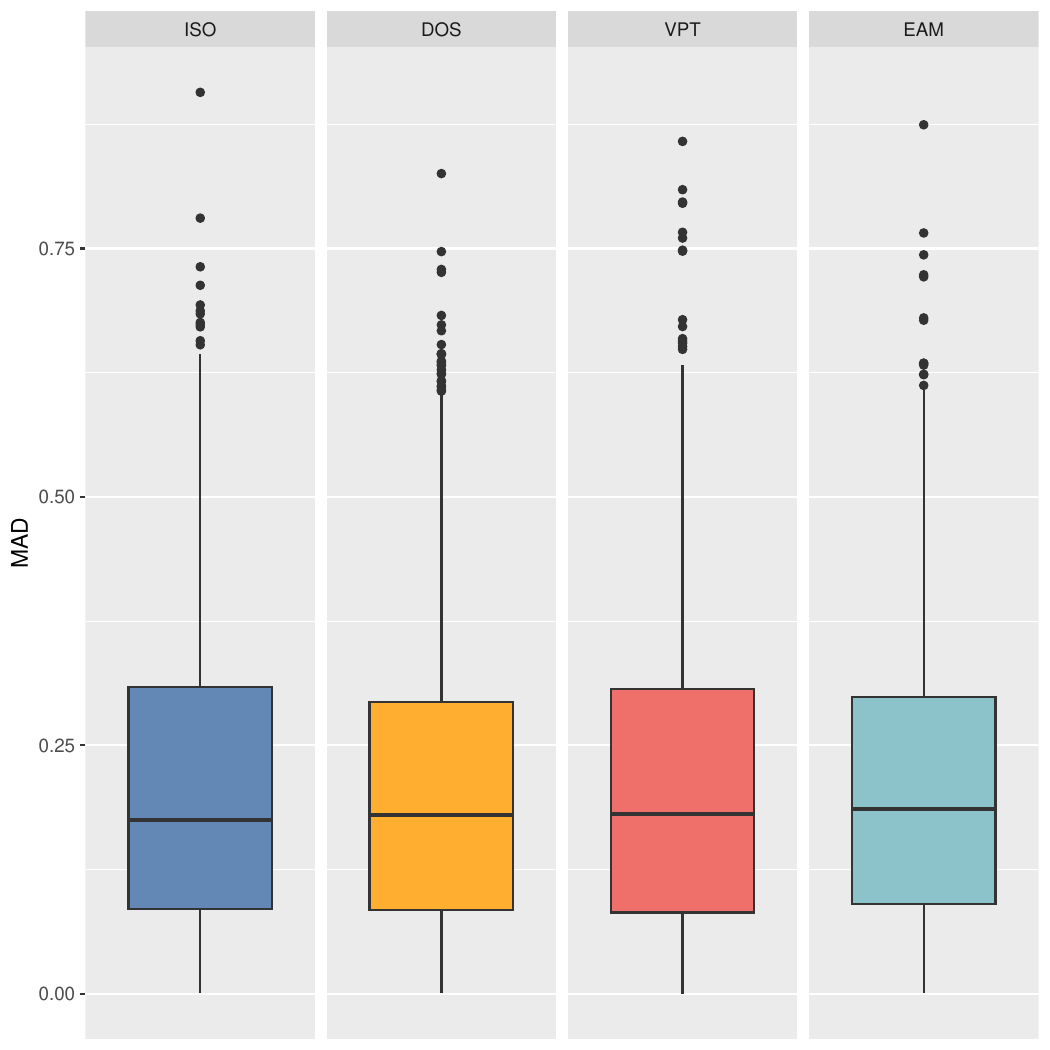}
    \caption{MAD with respect to the rendering method.}
    \label{fig:errors-method}
\end{figure}

First, we measured MAD by grouping the trials by method and volume. When grouping the trials by method (see \Cref{fig:errors-method}), the ISO rendering method exhibited the lowest MAD (MAD = 0.175, SD = 0.153), followed by DOS (MAD = 0.180, SD = 0.152), VPT (MAD = 0.181, SD = 0.156), and EAM (MAD = 0.186, SD = 0.151). We tested the null hypothesis that the MAD was equal between different methods. The normality and homoscedasticity assumptions were tested with Bartlett's and Shapiro-Wilk's tests, respectively. The normality assumption was unmet; therefore, we used the Kruskal-Wallis test. The test showed no significant differences in MAD between the methods ($H(3) = 0.85406, p = 0.8365$). Furthermore, we found no statistically significant bias of the mean depth difference.

\begin{figure}
    \centering
    \includegraphics[width=\linewidth]{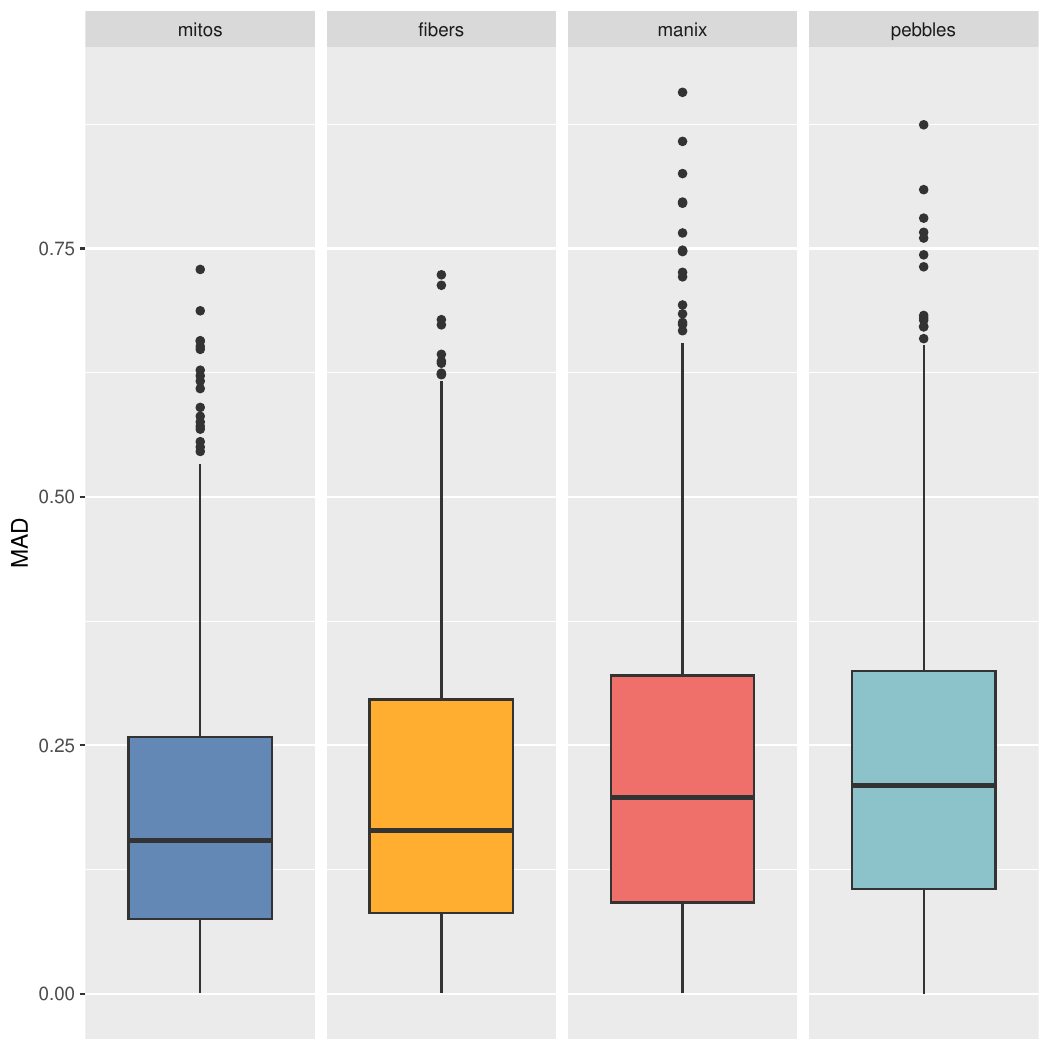}
    \caption{MAD with respect to the test volume.}
    \label{fig:errors-volume}
\end{figure}

When grouping the trials by volume (see \Cref{fig:errors-volume}), the lowest MAD was achieved on the mitos volume (MAD = 0.154, SD = 0.135), followed by fibers (MAD = 0.164, SD = 0.151), manix (MAD = 0.198, SD = 0.163), and pebbles (MAD = 0.210, SD = 0.157). We tested the null hypothesis that the MAD was equal between different volumes. The normality and homoscedasticity assumptions were tested with Bartlett's and Shapiro-Wilk's tests, respectively, and have not been met. Therefore, we used the Kruskal-Wallis test. The test showed a significant difference in MAD between the volumes ($H(3) = 55.262, p < 0.001$). Furthermore, Dunn's post-hoc test with Holm's adjustment showed significant differences between all pairs of volumes ($p < 0.05$) except between the manix volume and the pebbles volume. Furthermore, we found no statistically significant bias of the mean depth difference.

In order to address the potentially overbearing influence of test volumes when testing for differences in MAD between rendering methods, we additionally ran the Kruskal-Wallis test for each volume separately. Once more, our analysis revealed no statistically significant differences in MAD between the evaluated methods.

\begin{figure}
    \centering
    \includegraphics[width=\linewidth]{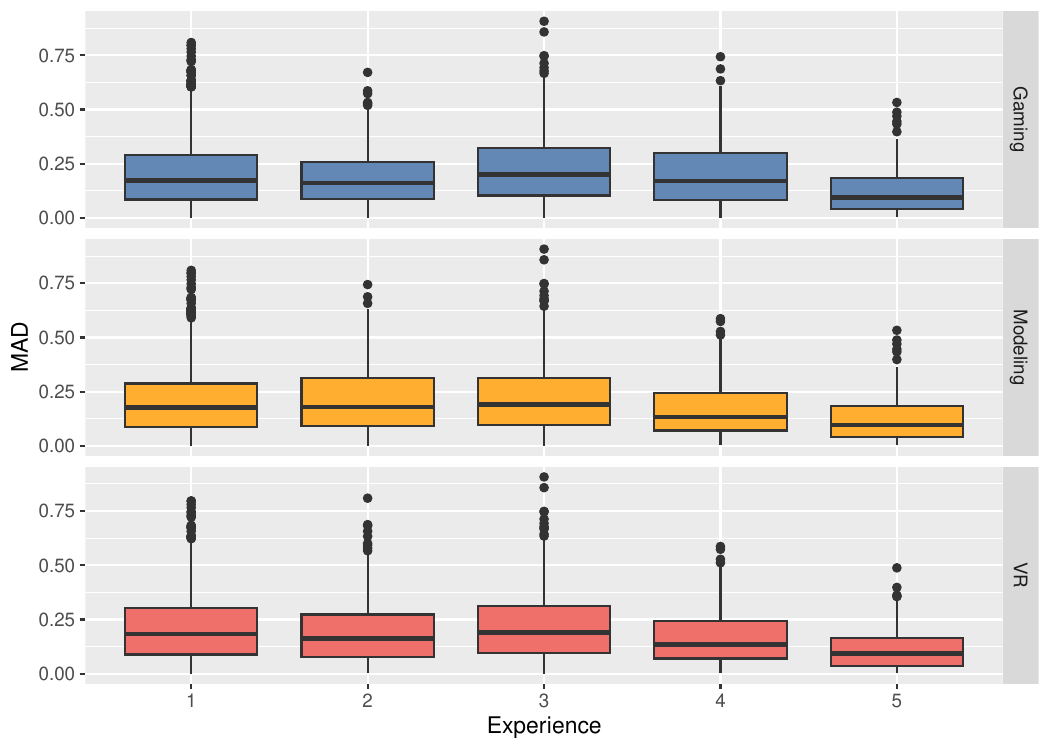}
    \caption{MAD with respect to the participants' prior experience with gaming, 3D modeling, and VR.}
    \label{fig:errors-experience}
\end{figure}

Next, we measured MAD by grouping the trials by the participants' prior experience with 3D environments (based on the participants who provided these data). A boxplot of the data is shown in \Cref{fig:errors-experience}. Since prior experience is a discrete variable expressed on a Likert scale, we used the Kruskal-Wallis test on a null hypothesis, stating that the MAD was equal between different experience levels. The test showed significant differences in all three experience categories: gaming ($H(4) = 64.84, p < 0.001$), 3D modeling ($H(4) = 59.12, p < 0.001$), and VR ($H(4) = 65.02, p < 0.001$). We used Dunn's post-hoc test with Holm's adjustment to find the experience levels that exhibited significant differences in MAD. In all three experience categories, the test showed that the participants who reported substantial experience (level 5 on the Likert scale) performed significantly better ($p < 0.001$) than those who reported less experience (levels 1-4 on the Likert scale). We found similar significant differences between experience levels 2 and 3, but we believe that this result might be skewed by psychological factors other than experience (see \Cref{sec:discussion}). Furthermore, we found weak negative correlations between MAD and experience levels in the VR category ($r = -0.01, F(1, 2957) = 16.77, p < 0.001$) and the 3D modeling category ($r = -0.006, F(1, 2957) = 7.30, p < 0.01$). We found no statistically significant correlation between MAD and age of the participant.

Additionally, we found statistically significant correlations between MAD and the positions of the bounding points. We found a positive correlation between MAD and 2D distance between bounding points on the screen ($r = 0.14, F(1, 4338) = 43.52, p < 0.001$), a slight negative correlation between MAD and 3D distance between bounding points in the volume ($r = -0.07, F(1, 4338) = 29.65, p < 0.001$), and a negative correlation between MAD and depth difference between bounding points in the volume ($r = -0.26, F(1, 4338) = 51.75, p < 0.001$).

%% file: content/05-discussion.tex
\section{Discussion}
\label{sec:discussion}

\begin{figure*}
    \centering
    \includegraphics[width=\linewidth]{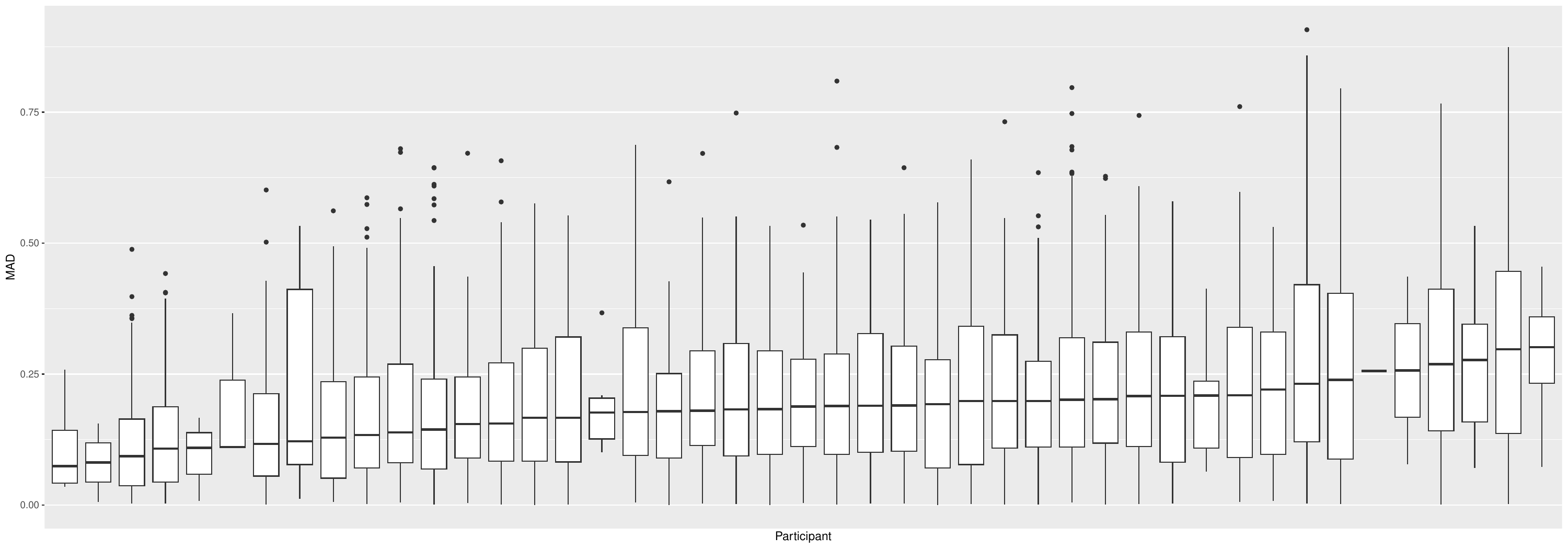}
    \caption{MAD for each participant. Note that not all participants evaluated all test cases.}
    \label{fig:errors-session}
\end{figure*}

The results of our study highlight several important aspects of participants' performance in a depth estimation task within crowded volumes. All participants achieved a MAD lower than the theoretical $1 / 3$ derived from uniformly random distance sampling and estimation. In general, MAD ranged consistently between 0.15 and 0.25 (see \Cref{fig:errors-session}).

Our analysis revealed no significant differences in MAD when grouping across different rendering methods (\Cref{fig:errors-method}). Traditional methods (EAM and ISO) yielded a MAD similar to that of modern, physically-based methods (DOS and VPT), which supports hypothesis (H3). This outcome aligns with the claims of several papers, which state that monocular cues resulting from camera movement, such as motion parallax, tend to outweigh the cues from shading and illumination~\cite{Boucheny2009,Englund2016,Englund2018}. In practical scenarios, where users typically control camera movement through direct interaction, we expect this effect to be even more pronounced, given the close connection between the human visual and motor systems~\cite{Wexler2005,Drouin2020}. We believe that the enduring popularity of traditional volume rendering methods, which has persisted for decades since their introduction, may be partly attributed to the influence of interaction.

\begin{figure*}
    \centering
    \includegraphics[width=\linewidth]{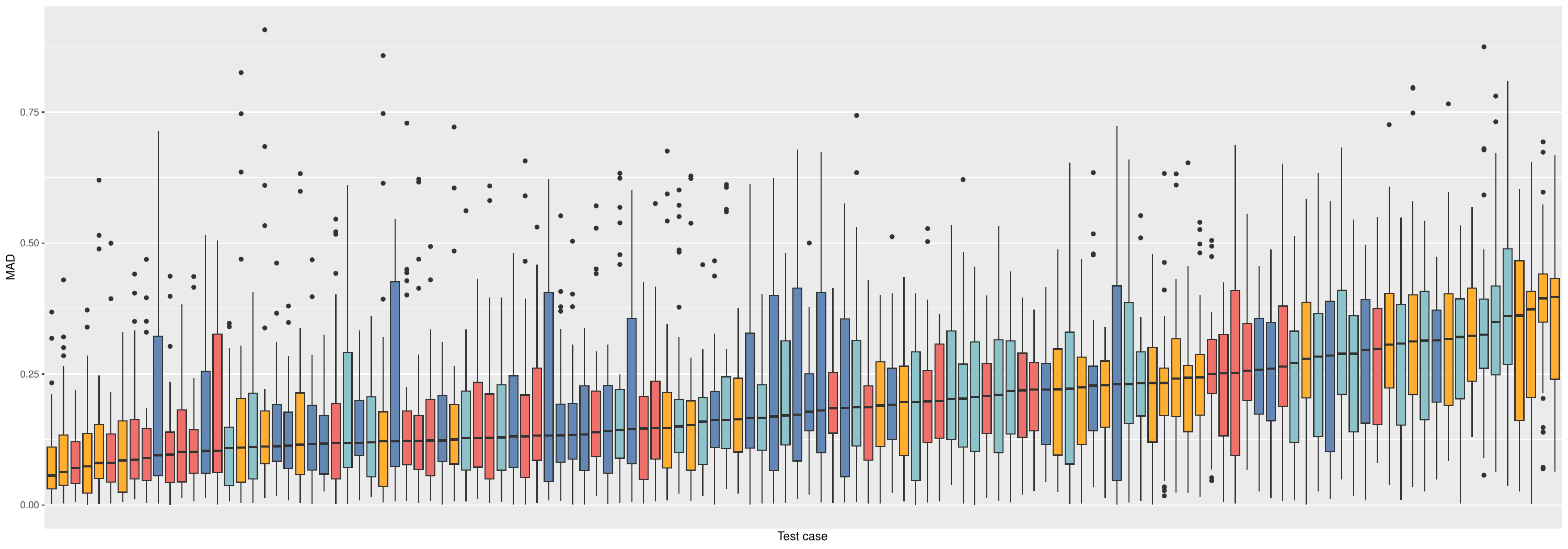}
    \caption{MAD for each test case. Color represents the volume used in the test case.}
    \label{fig:errors-video}
\end{figure*}

We found significant differences in MAD when we grouped across different volumes (\Cref{fig:errors-volume}), suggesting that the content of a volume has a greater impact on depth perception than the rendering method used for visualization. Interestingly, the manix volume, which does not exhibit crowdedness, ranked third among the evaluated volumes, with only the pebbles volume achieving a worse MAD. Familiarity with the shapes of human body parts does not appear to have contributed to a better score. The lower ranking may also be due to the transparency causing confusion about which structures were highlighted by the on-screen markers. Results from grouping across individual test cases (\Cref{fig:errors-video}) show that the manix volume ranked among the best and worst in terms of MAD, depending on the camera view. This suggests that certain camera views might introduce bias due to familiarity with the internal structures. In contrast, crowded volumes showed more consistent MAD values between camera views, indicating the absence of familiar features within these volumes.

The two top-ranked volumes were mitos and fibers, which is consistent with hypotheses (H1) and (H2). We attribute this ranking to the larger size of the structures they contain, leading to increased occlusion between them. Notably, the mitos volume achieved a lower MAD than the fibers volume, possibly due to the larger surface area of the structures, which made them more easily visible and identifiable within the crowded volume. This provides compelling evidence in support of hypothesis (H2). Conversely, the pebbles volume attained the highest MAD due to its substantial crowdedness and the small average size of the structures, which made it considerably more challenging to identify and visually track structures during camera movement. Participants informally confirmed this difficulty during post-experiment discussions. Indeed, when a structure is small, its visibility essentially becomes binary, eliminating partial occlusion as one of the strongest monocular depth cues~\cite{Englund2016} from the visualization. Furthermore, small structures do not significantly contribute to the overall illumination of the scene. They often produce small and soft shadows that have been shown to be confusing and detrimental to depth perception~\cite{Lindemann2011}. In accordance with (H1), this can be overcome by strong sparsification to reduce the number of visible instances and by introducing specific anchoring objects to create harder shadows.

Our analysis revealed significant correlations between MAD and reported experience with 3D environments. Specifically, participants with the highest reported experience levels consistently achieved higher accuracy in their depth estimates. This might stem from familiarity with 3D environments in general or with common biases inherent in 3D rendering, such as size bias~\cite{Boucheny2009}, depth bias~\cite{Howard2002,Diaz2017}, or shading bias~\cite{Mamassian1996,Langer2000}. Setting aside the highest experience levels, the slight negative correlation between MAD and experience level aligns with our expectation that prior experience with 3D environments positively influences depth perception. This is most evident when examining the performance of participants who reported no prior experience, as they consistently achieved worse MAD than more experienced participants. Additionally, we observed a significant difference in MAD between groups of participants that reported experience levels 2 and 3, which is likely related to the larger number of participants in the latter group. It is possible that participants with uncertain knowledge about their prior experience simply selected the middle option (level 3) instead of indicating no experience (level 1). Furthermore, our findings indicate that the age of the participants had no significant impact on performance. We had expected that older participants would either perform better due to their potentially greater experience with 3D images, or worse due to slower and less precise interaction with the computer and potential vision issues. The lack of age-related effects could be attributed to the absence of time constraints during the experiment that could have allowed for longer response times, or the artificial nature of the task itself, which would nullify any additional experience.

The analysis of correlations between MAD and the positions of the bounding points yielded the expected results. Participants achieved better performance when the bounding points were closely positioned on the screen. In contrast, focusing on two distant areas of the screen simultaneously means that the participants probably had to alternate their focus, which is difficult at best. During post-experiment discussions, participants noted that they often lost track of the bounding points during the initial animation, which made depth estimation considerably more difficult. The ability to replay the video slightly alleviated the issue. Furthermore, participants performed better when the depths of the bounding points differed substantially. This could result from a more pronounced motion parallax effect. Additionally, we believe that given a larger depth difference between the bounding points, participants could have fine-tuned their answers since they could leverage the contextual information provided by a larger neighborhood of structures surrounding the bounding points and the estimation point.

%% file: content/06-conclusion.tex
\section{Conclusion}
\label{sec:conclusion}

Depth perception is a challenging trait to measure, as it depends on the user's prior experience, environment, lighting setup, interaction, and the data itself. Every experiment aiming to measure depth perception must, therefore, clearly and precisely define all the above parameters. In contrast to previous studies, our study emphasizes the significance of camera movement as one of the most important features of visualization. While previous studies identified significant differences in user performance between different rendering methods with static images, our results demonstrate that even traditional methods without advanced lighting can perform as well as modern methods when camera movement is included. This provides strong evidence supporting hypothesis (H3).

Additionally, we showed that the volume's content affects depth perception more than the rendering method, which is in line with previous research~\cite{Lindemann2011,Grosset2013,Preim2016}. Most importantly, we used crowded volumes where we expected no familiarity with the structures contained in the volumes. The degree of crowdedness appears to negatively impact user performance due to the excessive amount of detail in the images, which is consistent with previous research~\cite{Lindemann2011,Preim2016} and hypothesis (H1). We showed that instances with a large screen surface area exhibit complex mutual occlusion, which positively affects depth perception since in this case the visibility of a single instance is not binary. This outcome is in line with hypothesis (H2). The worst-case scenario seems to be an extremely crowded volume with small instances, such as the pebbles volume in our study, where user performance was considerably worse than with other volumes.

Furthermore, we discovered correlations between user performance and the on-screen distance between the bounding points, and also between the depth difference between the bounding points. This finding underscores the importance of direct interaction, especially when the user can roughly align the structures of interest with the viewing direction. Such alignment minimizes the on-screen distance and maximizes the depth difference, allowing for better utilization of spatial context (the neighborhood of the structures of interest) and motion parallax.

Since our study was conducted on the general population, prior experience with 3D environments was found to be a contributing factor to the performance in the experiment. As expected, we observed a positive correlation between accuracy of depth estimation and the reported amount of prior experience.

The main limitation of our study is, of course, the lack of direct interaction between the participant and the visualization. This includes adjusting the camera view, the degree of sparsification, and the grouping and coloring of instances in the crowded environment. The reason for not including these aspects in the study is that the results would not be comparable between participants, as each participant interacts with the visualization in their own way. This opens up many possibilities for future research, especially in the context of direct interaction.